\documentclass[12pt,a4]{article}

\usepackage{amsmath}
\usepackage{amssymb}
\usepackage{url}
\usepackage{natbib}
\usepackage{siunitx}
\usepackage{setspace}

\usepackage[labelformat=simple]{subcaption}

\usepackage{graphicx}
\usepackage{color}

\usepackage{fancyvrb}

\usepackage{tikz}
\usetikzlibrary{positioning, arrows.meta, shapes.geometric}
\tikzset{%
  dot/.style n args = {4}{name=#3, circle, draw, inner sep=1pt, minimum size=5pt, fill=black, label={[shift={(#1,#2)}]#4:$#3$}},
  lat/.style n args = {4}{name=#3, circle, draw, inner sep=1pt, minimum size=5pt, label={[shift={(#1,#2)}]#4:$#3$}},
  sb/.style n args = {4}{name=#3, circle, draw, inner sep=1pt, minimum size=7pt, label={[shift={(#1,#2)}]#4:$#3$}},
  dot5/.style n args = {5}{name=#3, circle, draw, inner sep=1pt, minimum size=5pt, fill=black, label={[shift={(#1,#2)}]#4:$#5$}},
  lat5/.style n args = {5}{name=#3, circle, draw, inner sep=1pt, minimum size=5pt, label={[shift={(#1,#2)}]#4:$#5$}},
  sq/.style n args = {4}{name=#3, rectangle, draw, inner sep=1pt, minimum size=5pt, fill=black, label={[shift={(#1,#2)}]#4:$#3$}},
  tr/.style n args = {4}{name=#3, regular polygon,regular polygon sides=4, draw, inner sep=1pt, minimum size=6pt, fill=gray, label={[shift={(#1,#2)}]#4:$#3$}},
  bordered/.style = {draw,outer sep=1, inner sep=2, minimum size=5pt},
  >={Latex[width=1.5mm,length=2mm]},
  every picture/.style={semithick}
}

\definecolor{violet}{rgb}{0.7,0,0.7}
\definecolor{gray}{rgb}{0.4,0.4,0.4}

\newcommand{\cond}{\,\vert\,}
\newcommand{\given}{{ \, | \, }}

\newcommand{\E}{{\bf E}}

\newcommand{\NA}{\textrm{NA}}

\newcommand{\eref}[1]{(\ref{#1})}
\newcommand{\doo}{\textrm{do}}
\newcommand{\idsym}{Y}
\newcommand{\nonidsym}{--}

\newcommand{\citepp}[1]{\citep{#1}}

\begin{document}

\title{\vspace{-2.5cm}  Do-search -- a tool for causal inference and study design with multiple data sources}

\author{Juha Karvanen\\
Department of Mathematics and Statistics\\
University of Jyvaskyla, Finland\\
~\\
Santtu Tikka\\
Department of Mathematics and Statistics\\
University of Jyvaskyla, Finland\\
~\\
Antti Hyttinen\\
HIIT, Department of Computer Science\\
University of Helsinki, Finland
}

\date{}




\maketitle

\begin{abstract}
Epidemiological evidence is based on multiple data sources including clinical trials, cohort studies, surveys, registries and expert opinions. Merging information from different sources opens up new possibilities for the estimation of causal effects. We show how causal effects can be identified and estimated by combining experiments and observations in real and realistic scenarios. 
As a new tool, we present do-search, a recently developed algorithmic approach that can determine the identifiability of a causal effect. The approach is based on do-calculus, and it can utilize data with non-trivial missing data and selection bias mechanisms. When the effect is identifiable, do-search outputs an identifying formula on which numerical estimation can be based.
When the effect is not identifiable, we can use do-search to recognize additional data sources and assumptions that would make the effect identifiable. Throughout the paper, we consider the effect of salt-adding behavior on blood pressure mediated by the salt intake as an example. 
The identifiability of this effect is resolved in various scenarios with different assumptions on confounding.  
There are scenarios where the causal effect is identifiable from a chain of experiments but not from survey data, as well as scenarios where the opposite is true. As an illustration, we use survey data from NHANES 2013--2016 and the results from a meta-analysis of randomized controlled trials and estimate the reduction in average systolic blood pressure under an intervention where the use of table salt is discontinued. 


\end{abstract}

{\let\thefootnote\relax\footnote{{This is a preprint of the article that will appear in \textit{Epidemiology}.}}}

\section{Introduction}\label{sec:intro}

Epidemiological knowledge consists of cumulative evidence on associations and causal relations between background variables, risk factors and disease events. 
Traditional meta-analysis is commonly used to merge information from studies that are sufficiently similar according to predefined inclusion criteria \citepp{CochraneHandbook}. In a wider perspective, we may have a heterogeneous collection of studies available and the question is to decide whether these studies together allow for the causal effect of interest to be identified.  

For instance, consider identification of causal effect of $X$ on $Y$, defined here as the post-interventional \citepp{Pearl:book2009} distribution $P(Y \given \doo(X))$, in a setting where the effect  is mediated through $Z$. If there is an unobserved confounder between $X$ and $Z$, the causal effect is not identifiable from survey data on $X$, $Z$ and $Y$. Now assume that we carry out a new experiment  where $X$ is intervened on and $Z$ (but not $Y$) is measured. 
 By applying do-calculus \citepp{pearl1995causal} 
 we can show that the survey and the experiment together make it possible to identify the causal effect of $X$ on $Y$ (see Section~\ref{sec:variants} for details).

More generally, combining different data sources in a systematic way may be a challenging task. One has to perceive which variables are shared between the data sources, be aware of context-specific differences between the sources, understand the study design and missing data pattern, and recognize potential confounders. Graphical models can help to describe this information in an organized manner \citepp{textor1,Karvanen2015studydesign,textor2,matthay2020graphical}. Thereafter it remains to conclude whether the effect of interest is identifiable from the available data sources under the specified causal assumptions and if the answer is positive, to estimate the effect. 

A theoretical overview of recent developments in this kind of \emph{data fusion} is presented by  \citet{bareinboim2016pnas}.
The practical examples include propensity score methods for merging observational and experimental data \citepp{tipton2013improving,omuircheartaigh2014generalizing,rosenman2018propensity}, methods for causal inference in randomized controlled trials (RCTs) nested within cohorts of trial eligible individuals \citepp{dahabreh2018generalizing}, and average treatment effect estimation for pulmonary artery catheterization combining experimental with observational studies \citepp{hartman2015sate}. However, these examples do not address the general problem of deciding whether a causal effect can be identified from the available collection of experiments and observational studies in the presence of selection bias and missing data.

In this paper, we show how a recently developed algorithmic approach based on do-calculus, called \texttt{do-search} \citepp{dosearch,dosearchR}, can be applied in the epidemiological practice. \texttt{Do-search} can determine the identifiability of a causal effect when multiple data sources available. 
 The data sources may be observational or experimental and they may suffer from missing data and selection bias. \texttt{Do-search} can be used to derive 
expressions for causal effects and can be utilized in epidemiological research in guiding top-level study design,  evaluating consequences of missing data and merging information. We present examples that illustrate the use of \texttt{do-search} R package \citepp{dosearchR} as a part of the process of causal effect estimation.

As an illustration, 
we consider the causal effect of salt-adding behavior ($X$) on blood pressure ($Y$). Salt-adding behavior consists of the habits of adding salt in cooking and at the table. The salt intake ($Z$) is the mediator for the effect. 
In many countries, public health recommendations advise reducing the salt intake from the typical 9--12~g/day to 5--6~g/day \citepp{who2003diet,he2013effect}. 
While the salt intake may be difficult to measure in daily life, avoiding to add salt in cooking and at the table is a simple way to reduce it.
The effect of salt intake on blood pressure have been established in many studies, and several reviews and meta-analyses have been carried out\citepp{he2002effect,he2009comprehensive,graudal2012effects,aburto2013effect,he2013effect}. 
For instance, on the basis of a meta-analysis of 34 RCTs (3230 participants),  a \SI{100}{mmol} (\SI{6}{g}) reduction in 24 hour urinary sodium was associated with a fall in systolic blood pressure of \SI{5.8}{mm Hg} after adjustment for age, ethnic group, and blood pressure status \citepp{he2013effect}. In addition to salt intake, other dietary and life style factors as well as genetic factors are known to affect blood pressure \citepp{poulter2015hypertension}. Many of these factors may also be associated with low salt preference and salt-adding behavior. 

The question of interest is to find when the causal effect of salt-adding behavior ($X$) on blood pressure ($Y$) can be estimated without a direct experiment where $X$ is intervened and $Y$ is measured.
We will consider different causal structures and different combinations of data sources and show how \texttt{do-search} can be applied to determine the identifiability of the causal effect in these scenarios. As a real data illustration we combine the meta-analytical results mentioned above\citepp{he2013effect}, and observational data from the National Health and Nutrition Examination Survey (NHANES) 2013--2016 surveys.

\section{Methods}
\subsection{Concepts and Notation} \label{sec:concepts}
Expert knowledge on the causal mechanisms is an essential element of causal inference, and a causal model is a way to formalize this knowledge. A structural causal model \citepp{Pearl:book2009} specifies 
the known or hypothesized causal relations between a set of variables. These relations are often represented by directed acyclic graphs (DAGs) or their semi-Markovian extensions where latent common causes are marked by bidirected arcs between two observed variables  \citep{Shpitser} (see Figure~\ref{fig:frontdoor} for an example). 

An intervened variable $X$ is marked with the do-operator as $\doo(X)$ and the causal effect  $P(Y \cond \doo(X))$ denotes the distribution of $Y$ when $X$ is forced to a value by the intervention. 
A causal effect is identifiable if it can be uniquely determined from the known distributions. When an identifiable causal effect is estimated, these distributions are usually replaced by their parametric or nonparametric estimates. 
 In simple cases, identifiability can be checked by manually applying standard probability calculus and do-calculus \citepp{pearl1995causal}. Do-calculus consists of rules for inserting and deleting observations, exchanging observations and interventions, and inserting and deleting interventions.  
There exists efficient algorithms for determining identifiability for settings where data from a single observational source \citepp{Shpitser}, from multiple domains \citepp{bareinboim2013general}, or from surrogate experiments \citepp{Bareinboim:zidentifiability,Tikka:surrogate,lee2019general} are available. An open-source software implementation for many of these algorithms are available as well \citepp{Tikka:identifying}.

\subsection{Data Sources}

In the general setup, the available data sources (inputs) include multiple observational and experimental studies whose respective distributions can be described in a symbolic form, i.e., as an expression such as 
``$P(X,Z,Y)$'' or ``$P(Z \cond \doo(X))$''. Now a causal effect is identifiable if it can be uniquely expressed as a formula using only the inputs and quantities derivable from them. For instance, $P(X)$ is directly derivable from $P(X,Z,Y)$ and $P(Z \cond \doo(X), W)$ is derivable from $P(Z \cond \doo(X))$ if the conditional independence of $Z$ and $W$ given $X$ is implied in the graph where the incoming edges to $X$ are removed. The rules of do-calculus are valid  under the general setup but the algorithms \citepp{Shpitser,bareinboim2013general,Bareinboim:zidentifiability,Tikka:surrogate,lee2019general} mentioned in Section~\ref{sec:concepts} only work in special cases.  

Two sources of data are used for a numeric illustration. The meta-analysis of 34 RCTs \citepp{he2013effect} provides information on the causal effect of salt intake ($Z$) on blood pressure ($Y$). We use the summary of the original trials as our input data. The summary reports the mean change in urinary sodium during the study (in mmol/24h) and the mean change in systolic blood pressure (in mm Hg) for each study. The available study-level background variables ($W$) include the mean age, the proportion of males, the proportion of white people and hypertension status (hypertensive or normotensive). The data source is written symbolically as $P(Y \cond \doo(Z), W)$.

NHANES 2013--2016 questionnaire data provide information on the salt intake ($Z$) and salt-adding behavior ($X$) in the United States.  The participants have recorded the dietary items they have consumed on two days and the daily sodium intake has been derived from these items. We use the mean of two sodium measurements. The different units are transformed using the equality that \SI{100}{mmol} of salt (NaCl) weights \SI{5.844}{g} and contains \SI{2.299}{g} (\SI{39.34}{mmol}) of sodium (Na). In order to measure the salt-adding behavior we derive a salt score that consists of three questions: 
\begin{enumerate}
 \item \textit{How often do you add ordinary salt to your food at the table?} (Rarely 0, Occasionally 1, Very often 2)
 \item \textit{Did you add any salt to your food at the table yesterday?} (No 0, Yes 1), and
 \item \textit{How often is ordinary salt or seasoned salt added in cooking or preparing foods in your household?} (Never 0, Rarely 1, Occasionally 2, Very often 3).
\end{enumerate}
The salt score is the sum of the values of these three questions and attains values from 0 to 6. 
The common background variables, age, gender, ethnicity (white or non-white) and hypertension status (hypertensive or  normotensive), are the same variables $W$ as in the meta-analysis. The additional background variables ($H$) include education and the eating out frequency (times per month). This data source is written symbolically as $P(X,Z,W,H)$.
The basic demographic variables have been collected for 11488 individuals and the analysis data set contains 9957 individuals who have the salt score, the eating out frequency (times per month) and at least one measurement of sodium intake available.  The sampling weights provided with the data are used in all analyses.  

\subsection{Examining Identifiability from Multiple Data Sources with Do-search}\label{sec:dosearch}

\texttt{Do-search} is an open-source software that has been designed for non-parametric identification problems when multiple data sources are available \citepp{dosearch,dosearchR}. The algorithm aims to derive the causal effect of interest from the inputs by carrying out a systematic search over the rules of do-calculus and marginalization, conditioning and chain rule multiplication permitted by probability calculus. The algorithm derives new identifiable distributions by applying these rules to the distributions that have been given as the input or have been identified in the previous steps. This process is repeated until the algorithm encounters the target distribution or cannot identify any new distributions.
Note that \texttt{do-search} is not related to causal search algorithms in causal discovery \citepp{glymour2019review}.  Since the approach is search-based, the computational load increases rapidly when the number of variables grows. This can be often mitigated by grouping similar variables in the graph. \texttt{Do-search} uses heuristics and search-space reduction techniques that speed up the algorithm in the vast majority of cases. 

The formulas returned by \texttt{do-search} are fully non-parametric. The representation of the formulas  assumes that the variables are discrete but the summations can be changed to integrals if the corresponding variables are continuous. Given an identifying formula, the estimation of the causal effect is a statistical problem for which the full repertoire of statistical and machine learning methods is available. 

\texttt{Do-search} can also cope with missing data problems. The graph is augmented by adding nodes for  measurements and response indicators that specify whether the value of the variable is measured or not \citepp{Mohan2013,Karvanen2015studydesign}. A measurement $X^*$ is linked to the true value $X$ and response indicator $R_X$ as follows
\begin{equation} \label{eq:missingness}
X^* = 
\begin{cases}
  X , & \textrm{if}\; R_{X} = 1, \\
  \NA,  & \textrm{if}\; R_{X} = 0,
\end{cases}
\end{equation}
where $\NA$ denotes a missing value. For instance, the input $P(X^*,Z^*,Y^*,R_X,R_Z,R_Y)$ refers to an observational study where variables $X$, $Z$ and $Y$ suffer from missing data. When missing data are present, \texttt{do-search} uses additional inference rules to take response indicators into account\citepp{dosearch}. These rules are not directly related to recent theoretical work on identification under missing data \citepp{Mohan2013,Shpitser2015,bhattacharya2019}.

\texttt{Do-search} is sound, meaning that formulas produced for queries found to be identifiable by the algorithm are always correct. 
Although the rules of the search have been shown to be complete in several restricted problem settings \citepp{dosearch,Shpitser,bareinboim2013general,Bareinboim:zidentifiability,lee2019general},
they have not been shown to completely characterize identifiability when the data come from multiple sources. 
In practice this means that if we wish to confirm that a causal effect is not identifiable
we need to resort to further study of the specific problem to rule
out the possibility that an identifying formula could be derived by some other means.

\section{Results}
\subsection{The Front-Door Setting with Multiple Data Sources} \label{sec:frontdoor}

\begin{figure}
\begin{center}

\begin{subfigure}[t]{1.0\textwidth}
 \centering
 \begin{tikzpicture}[scale=2.0]
\node [dot = {-0.05}{0}{X}{below}] at (0,0) {};
\node [dot = {0}{0}{Z}{below}] at (1,0) {};
\node [dot = {0}{0}{Y}{below}] at (2,0) {};

\draw [->] (X) -- (Z);
\draw [->] (Z) -- (Y);
\draw [<->,dashed] (X) to [bend left=40]  (Y);
\end{tikzpicture}
\caption{Basic front-door.} \label{fig:basicfrontdoor}
\end{subfigure}

\begin{subfigure}[t]{0.465\textwidth}
 \centering
 \begin{tikzpicture}[scale=2.0]
\node [dot = {-0.05}{0}{X}{below}] at (0,0) {};
\node [dot = {0}{0}{Z}{below}] at (1,0) {};
\node [dot = {0}{0}{Y}{below}] at (2,0) {};

\draw [->] (X) -- (Z);
\draw [->] (Z) -- (Y);
\draw [<->,dashed] (X) to [bend left=40]  (Z);
\end{tikzpicture}
\caption{Unobserved pre-mediator confounding.} \label{fig:bidirectedXZ}
\end{subfigure}
\hfill
\begin{subfigure}[t]{0.475\textwidth}
 \centering
\begin{tikzpicture}[scale=2.0]
\node [dot = {-0.05}{0}{X}{below}] at (0,0) {};
\node [dot = {0}{0}{Z}{below}] at (1,0) {};
\node [dot = {0}{0}{Y}{below}] at (2,0) {};

\draw [->] (X) -- (Z);
\draw [->] (Z) -- (Y);
\draw [<->,dashed] (X) to [bend left=45]  (Y);
\draw [<->,dashed] (X) to [bend right=35]  (Z);
\end{tikzpicture}  
\caption{Front-door with unobserved pre-mediator confounding.} \label{fig:frontdoorwithXZ}
\end{subfigure}

\begin{subfigure}[t]{0.465\textwidth}
 \centering
\begin{tikzpicture}[scale=2.0]
\node [dot = {0}{0}{X}{below}] at (0,0) {};
\node [dot = {-0.05}{0}{Z}{below}] at (1,0) {};
\node [dot = {0.05}{0}{Y}{below}] at (2,0) {};

\draw [->] (X) -- (Z);
\draw [->] (Z) -- (Y);
\draw [<->,dashed] (Z) to [bend left=35]  (Y);
\end{tikzpicture}
\caption{Unobserved post-mediator confounding. } \label{fig:bidirectedZY}
\end{subfigure}
\hfill
\begin{subfigure}[t]{0.475\textwidth}
 \centering
\begin{tikzpicture}[scale=2.0]
\node [dot = {0}{0}{X}{below}] at (0,0) {};
\node [dot = {-0.05}{0}{Z}{below}] at (1,0) {};
\node [dot = {0.05}{0}{Y}{below}] at (2,0) {};

\draw [->] (X) -- (Z);
\draw [->] (Z) -- (Y);
\draw [<->,dashed] (X) to [bend left=45]  (Y);
\draw [<->,dashed] (Z) to [bend right=35]  (Y);
\end{tikzpicture}
\caption{Front-door with unobserved post-mediator confounding. }   \label{fig:frontdoorwithZY}
\end{subfigure}

\begin{subfigure}[t]{0.465\textwidth}
 \centering
 \begin{tikzpicture}[scale=2.0]
\node [dot = {-0.05}{0}{X}{below}] at (0,0) {};
\node [dot = {0}{0}{Z}{below}] at (1,0) {};
\node [dot = {0.05}{0}{Y}{below}] at (2,0) {};
\node [dot = {0}{0}{W}{above}] at (1,0.618) {};

\draw [->] (X) -- (Z);
\draw [->] (Z) -- (Y);
\draw [->] (W) -- (Y);
\draw [->] (W) -- (Z);
\draw [->] (W) -- (X);
\draw [<->,dashed] (X) to [bend right=35]  (Y);
\end{tikzpicture}  
\caption{Front-door with an observed confounder. }  \label{fig:frontdoorW}
\end{subfigure}
\hfill
\begin{subfigure}[t]{0.475\textwidth}
 \centering
\begin{tikzpicture}[scale=2.0]
\node [dot = {-0.05}{0}{X}{below}] at (0,0) {};
\node [dot = {0}{0}{Z}{below}] at (1,0) {};
\node [dot = {0.05}{0}{Y}{below}] at (2,0) {};
\node [dot = {0}{0}{W}{above}] at (1,0.618) {};

\draw [->] (X) -- (Z);
\draw [->] (Z) -- (Y);
\draw [->] (W) -- (Y);
\draw [->] (W) -- (Z);
\draw [->] (W) -- (X);
\draw [<->,dashed] (X) to [bend right=35]  (Y);
\draw [<->,dashed] (X) to [bend right=35]  (Z);
\end{tikzpicture}  
\caption{Front-door with an observed confounder and unobserved pre-mediator confounding. } \label{fig:frontdoorWwithXZ}
\end{subfigure}
\caption{Causal models where the causal effect of $X$ on $Y$ is mediated by $Z$. In the example, $X$ stands for salt-adding behavior, $Z$ for salt intake, $Y$ for blood pressure and $W$ for common confounders.} \label{fig:frontdoor}

\end{center}
\end{figure}
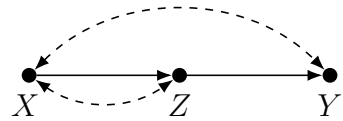
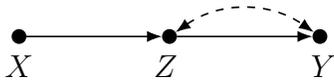
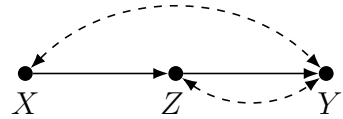
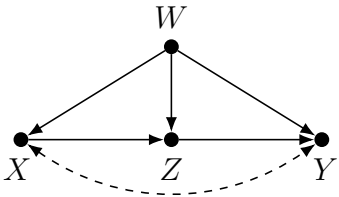
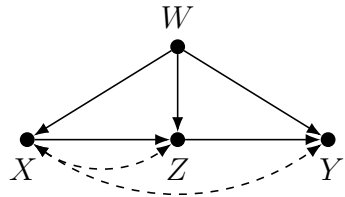

We study graphs where the effect of $X$ on $Y$ is mediated through $Z$ because this structure leads to many interesting scenarios. This is not a restriction for the approach but \texttt{do-search} is fully applicable also when the graph contains an edge from $X$ to $Y$ or some other graphical structure.  The well-know front-door setting \citepp{pearl1995causal} is shown in Figure~\ref{fig:basicfrontdoor}.  First, consider a scenario where salt-adding behavior, salt intake and blood pressure have been measured in a population-based survey. If the sample does not suffer from selection bias, we have data on the joint distribution $P(X,Z,Y)$. The causal effect can be identified by the front-door adjustment formula  
\[
 P(Y \given \doo(X)) = \sum_Z P(Z \given X) \sum_{X^\prime} P(X^\prime) P(Y \given X^\prime,Z),
\]
where all marginal and conditional distributions can be estimated from the survey data.

There are also other ways to identify $P(Y \cond \doo(X))$. Instead of data on $P(X,Z,Y)$, the available data sources  could include an experiment that provides information on $P(Y \cond \doo(Z))$ and a survey that provides information on $P(X,Z)$. Applying \texttt{do-search} we obtain
\begin{equation} \label{eq:surveyandexperiment}
 P(Y \given \doo(X)) = \sum_Z P(Z \given X) P(Y \given \doo(Z)),
\end{equation}
where the first term can be estimated from the survey and the second term from the experiment.
The R code for deriving this result with \texttt{do-search} is presented in Figure~\ref{fig:Rcode}. 

\begin{figure}
\begin{center}
\begin{minipage}{8cm}
\begin{Verbatim}[fontsize=\small, frame=single, framesep=2mm, baselinestretch=0.9]
library(dosearch)

graph <- "
    X -> Z
    Z -> Y
    X <-> Y"

data <- "
    P(Y | do(Z))
    P(X,Z)"

query <- "P(Y | do(X))"

dosearch(data, query, graph)

------------- Output --------------
$identifiable
[1] TRUE

$formula
[1] "[sum_{Z} [p(Z|X)*p(Y|do(Z))]]"
\end{Verbatim}
\end{minipage}
\end{center}
\caption{Example R code on the use \texttt{do-search} to determine identifiability of $P(Y \cond \doo(X))$ from $P(Y \cond \doo(Z))$ and $P(X,Z)$ under the assumptions encoded in the graph of Figure \ref{fig:basicfrontdoor}. The R codes for the other examples of this paper with \texttt{do-search} are available in Appendix A.  
}
\label{fig:Rcode}
\end{figure}

On the contrary, some other combinations of studies do not lead to identification in the graph of Figure~\ref{fig:basicfrontdoor}. 
For instance, the collection of three surveys providing information on $P(X,Z)$, $P(X,Y)$ and $P(Z,Y)$ and an experiment providing information on $P(Z \cond \doo(X))$ is not sufficient to identify $P(Y \cond \doo(X))$ (Formal proofs for non-identifiability are given in Appendix B). 

\subsection{Variants of the Front-Door Setting} \label{sec:variants}
Next we will study identifiability in variants of the basic front-door setting by utilizing \texttt{do-search}. The graphs for these settings are shown in Figure~\ref{fig:frontdoor} and the identifiability results for different data sources are summarized in  Table~\ref{tab:IDresults}. 

\begin{table}
\caption{Identifiability of $P(Y \cond \doo(X))$ from different data sources in the graphs of Figure~\ref{fig:frontdoor}. The data sources are characterized by the underlying theoretical distributions; e.g., a survey may provide information on $P(X,Y,Z)$ and an experiment may provide information on $P(Y \cond \doo(Z))$. Symbol \idsym{} denotes identifiability and symbol \nonidsym{} non-identifiability (indicated by \texttt{do-search} and proven in Appendix B). 
For graphs a--e of Figure~\ref{fig:frontdoor} that do not include $W$, we assume that $W$ is not connected to any other vertices of the graph.%
}
\label{tab:IDresults}
\begin{center}
\begin{tabular}{llllllllllll}
 & & \multicolumn{7}{c}{Graph of Figure~\ref{fig:frontdoor}} \\
&Data sources & a & b & c & d & e & f & g\\
\hline
1.& $P(X,Y,Z)$  &  \idsym  &  \nonidsym  &  \nonidsym  &  \idsym  &  \nonidsym  &  \nonidsym  &  \nonidsym \\ 
2.& $P(X,Z),   P(Y \cond \doo(Z))$  &  \idsym  &  \nonidsym  &  \nonidsym  &  \nonidsym  &  \nonidsym  &  \nonidsym  &  \nonidsym \\ 
3.& $P(Z \cond \doo(X)),   P(Y \cond \doo(Z))$  &  \idsym  &  \idsym  &  \idsym  &  \nonidsym  &  \nonidsym  &  \nonidsym  &  \nonidsym \\ 
4.& $P(Z,Y),   P(Z \cond \doo(X))$  &  \nonidsym  &  \idsym  &  \nonidsym  &  \nonidsym  &  \nonidsym  &  \nonidsym  &  \nonidsym \\ 
5.& $P(X,Z),   P(X,Y),   P(Z,Y),   P(Z \cond \doo(X))$  &  \nonidsym  &  \idsym  &  \nonidsym  &  \idsym  &  \nonidsym  &  \nonidsym  &  \nonidsym \\ 
6.& $P(X,Y,Z,W)$  &  \idsym  &  \nonidsym  &  \nonidsym  &  \idsym  &  \nonidsym  &  \idsym  &  \nonidsym \\ 
7.& $P(X,Z,W),   P(Y \cond \doo(Z),W)$  &  \idsym  &  \nonidsym  &  \nonidsym  &  \nonidsym  &  \nonidsym  &  \idsym  &  \nonidsym \\ 
8.& $P(Z \cond \doo(X),W),   P(Y \cond \doo(Z),W)$  &  \idsym  &  \idsym  &  \idsym  &  \nonidsym  &  \nonidsym  &  \nonidsym  &  \nonidsym \\ 
9.& $P(Z \cond \doo(X),W),   P(Y \cond \doo(Z),W),   P(W)$  &  \idsym  &  \idsym  &  \idsym  &  \nonidsym  &  \nonidsym  &  \idsym  &  \idsym 
\end{tabular}
\end{center}
\end{table}

Figure~\ref{fig:bidirectedXZ} shows the scenario described in Section~\ref{sec:intro}. The causal effect $P(Y \given \doo(X))$ is not identifiable from $P(X,Y,Z)$ (line 1 and column b in Table~\ref{tab:IDresults}). When an experiment providing information on $P(Z \cond \doo(X))$ and a survey providing information on $P(Z,Y)$ are available (line 4 of Table~\ref{tab:IDresults})  the causal effect can be identified: 
\begin{equation*} 
  P(Y \given \doo(X)) = \sum_Z P(Z \given \doo(X)) P(Y \given Z).
\end{equation*}

If there is unobserved pre-mediator confounding in the front-door setting (Figure~\ref{fig:frontdoorwithXZ}),
neither $P(X,Z,Y)$ nor $P(Y \cond \doo(Z))$ and $P(X,Z)$ are sufficient to identify $P(Y \given \doo(X))$ (lines 1--2). In this situation, a chain of experiments providing information on $P(Z \cond \doo(X))$ and  $P(Y \cond \doo(Z))$ (line 3) makes the causal effect identifiable: 
\begin{equation} \label{eq:chain_of_experiments}
 P(Y \given \doo(X)) = \sum_Z P(Z \given \doo(X)) P(Y \given \doo(Z)).
\end{equation}

If instead, we have post-mediator confounding like in Figure~\ref{fig:bidirectedZY} the situation changes and the chain of experiments (line 3) does not produce identifiability. 
An intuitive explanation for this can be given by looking at the structural equations under intervention $\doo(X=x)$:
\begin{align*}
 & X = x, \\
 & Z = f_Z(x,U), \\
 & Y = f_Y(Z,U).
\end{align*}
Here $U$ is the unobserved confounder that affects both $Z$ and $Y$. In two separate experiments, $P(Z \given \doo(X))$ and $P(Y \given \doo(Z))$, confounder $U$ is not shared between the experiments. For this reason, equation~\eref{eq:chain_of_experiments} does not specify a correct formula for $P(Y \given \doo(X))$ in this case.
As an extreme example, let $X$, $Z$, $Y$ and $U$ be binary and specify $P(U=1)=0.5$, $Z=X \oplus U$ and $Y=Z \oplus U$, where $\oplus$ stands for the exclusive logical disjunction. Now, since $P(Z \given \doo(X))=0.5$ and  $P(Y \given \doo(Z))=0.5$, equation~\eref{eq:chain_of_experiments} suggests $P(Y \given \doo(X)) = 0.5$ for any value of $X$ and $Y$. However, intervened value of $X$ perfectly determines $Y$.
Naturally, the causal effect of $X$ on $Y$ can be identified from $P(X,Y)$  (line 1) directly as $P(Y \given \doo(X)) = P(Y \given X)$. 

If post-mediator confounding occurs in the front-door setting (Figure~\ref{fig:frontdoorwithZY}), 
additional data sources such as $P(X,Z,Y)$ or $P(Y \given \doo(Z), X)$ do not help (lines 1--9) and  $P(Y \given \doo(X))$ is identifiable only from an experiment where $X$ is intervened and $Y$ is measured. 

Figures~\ref{fig:frontdoorW} and \ref{fig:frontdoorWwithXZ} present variants where covariate $W$ is observed. In Figure~\ref{fig:frontdoorW}, the causal effect $P(Y \given \doo(X))$ can be identified from $P(X,Z,Y,W)$ (line 6) as
\begin{equation*}
  \sum_{Z,W} P(W)P(Z \given X,W) \sum_{X^\prime} P(X^\prime \given W) P(Y \given X^\prime,Z,W)
\end{equation*}
or from the combination of an experiment providing information on $P(Y \given \doo(Z),W)$ and a survey providing information on $P(X,Z,W)$ (line 7) as
\begin{equation*}
 \sum_{Z,W} P(W)P(Z \given X,W)P(Y \given \doo(Z),W). 
\end{equation*}
However, the causal effect is not identifiable from a chain of experiments providing information on $P(Z \given \doo(X),W)$ and $P(Y \given \doo(Z),W)$ (line 8) unless the marginal distribution $P(W)$ is also known (line 9). In Figure~\ref{fig:frontdoorWwithXZ}, this combination of two experiments and a survey (line 9) allows the causal effect $P(Y \given \doo(X))$ to be identified by 
\begin{equation*}
 \sum_{Z,W} P(W)P(Z \given \doo(X),W)P(Y \given \doo(Z),W). 
\end{equation*}
A survey providing information on $P(X,Z,Y,W)$ (line 6) or the combination of an experiment providing information on $P(Y \given \doo(Z),W)$ and a survey providing information on $P(X,Z,W)$ (line 7) are not sufficient for identification in this case.

\subsection{Illustration with Real Data} \label{sec:realdataexample}

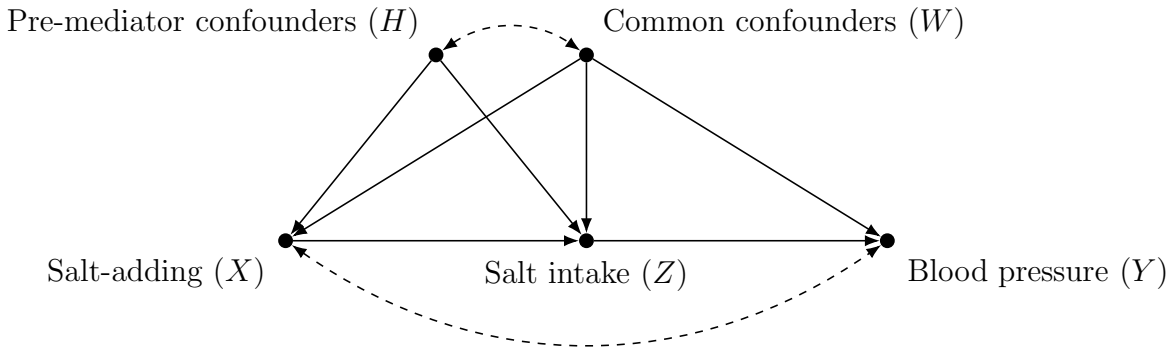
\begin{figure}
\begin{center}
 \begin{tikzpicture}[scale=4.0]
\node [dot5 = {-0.05}{0}{X}{below left}{\textrm{Salt-adding } (X)}] at (0,0) {};
\node [dot5 = {0}{0}{Z}{below}{\textrm{Salt intake } (Z)}] at (1,0) {};
\node [dot5 = {0.05}{0}{Y}{below right}{\textrm{Blood pressure } (Y)}] at (2,0) {};
\node [dot5 = {0}{0}{W}{above right}{\textrm{Common confounders } (W)}] at (1,0.618) {};
\node [dot5 = {0}{0}{H}{above left}{\textrm{Pre-mediator confounders } (H)}] at (0.5,0.618) {};

\draw [->] (X) -- (Z);
\draw [->] (Z) -- (Y);
\draw [->] (W) -- (Y);
\draw [->] (W) -- (Z);
\draw [->] (W) -- (X);
\draw [->] (H) -- (X);
\draw [->] (H) -- (Z);
\draw [<->,dashed] (X) to [bend right=35]  (Y); 
\draw [<->,dashed] (H) to [bend left=35]  (W); 
\end{tikzpicture}
\caption{Causal model for the real data illustration.} \label{fig:realdata}
\end{center}
\end{figure}

We aim to estimate the mean change in systolic blood pressure in the US population under an intervention that makes everyone avoid adding salt to their food (in preparation or at table). More technically, the intervention is defined as setting the salt score to zero for everyone in the NHANES 2013--2016 surveys. Figure~\ref{fig:realdata} presents a causal model for the situation. Recall that the NHANES data provides information on the observational distribution $P(X,Z,H,W)$ and the meta-analysis provides information on the experimental distribution $P(Y \given \doo(Z),W)$. The target to be estimated is the causal effect $P(Y \given \doo(X))$. Applying \texttt{do-search} we obtain
\begin{equation} \label{eq:realdataP}
 P(Y \given \doo(X)) = \sum_{Z,H,W} P(H,W) P(Z \given X,H,W) P(Y \given \doo(Z),W).
\end{equation}
As we fit a statistical model for the expected value of $Y$, we write equation~\eref{eq:realdataP} in the form where the distribution of $Y$ is replaced by expectation
\begin{equation} \label{eq:realdataE}
 \E(Y \given \doo(X)) = \sum_{Z,H,W} P(H,W) P(Z \given X,H,W) \E(Y \given \doo(Z),W).
\end{equation}
Formula~\eref{eq:realdataE} shows that three models are needed: a model for the joint distribution $P(H,W)$, a model that explains the salt intake $Z$ by $X$, $H$ and $W$, and a model that explains the systolic blood pressure $Y$ by $Z$ and $W$. The first model can be replaced by the empirical joint distribution of $H$ and $W$, i.e., calculating the average over the values in the data.  The second model is estimated from the NHANES 2013--2016 data. We fit a linear model for sodium intake with covariates salt score, gender, age, education and ethnicity (white or non-white). As formula~\eref{eq:realdataE} is non-parametric, the second model could be non-linear model as well. The estimated regression coefficients and their confidence intervals are shown in Table~\ref{tab:realdata_regression}. According to the model, the difference in the salt intake between salt score values 6 and 0 equals \SI{0.46}{g} (\SI{20.1}{mmol}) of sodium (\SI{1.2}{g} of salt). 

The third model is a meta-regression model where the change in systolic blood pressure is explained by the change in urinary sodium, hypertension status (hypertensive or normotensive), mean age, the proportion of males and the proportion of participants classified as ``white'', and the remaining heterogeneity between the studies is modeled by a random effect. 
The estimated regression coefficients and their confidence intervals estimated with the R package \texttt{metafor} \citepp{metafor} are shown in Table~\ref{tab:realdata_regression}. 

\begin{table} 
\caption{The estimated regression coefficients for the linear model explaining the sodium intake (in mg) and for the meta-regression model the explaining the change in systolic blood pressure (in mm Hg). CI stands for confidence intervals and UNa for urinary sodium.  
\label{tab:realdata_regression}} 
\begin{center}
\begin{tabular}{lc}
\multicolumn{2}{l}{\textbf{The model for the sodium intake} $\E(Z \given X,H,W)$}\\
&  \\
Parameter & Estimate (95\% CI) \\
\hline
 Intercept  &  ~~~2927.7    (2770.9,\,3084.6)~~~ \\ 
 Salt score (0--6)  &  77.4    (58.2,\,96.5)~~~ \\ 
 Eating out (times/month)  &  54.4    (42.7,\,66.2)~~~ \\ 
 Age (years)  &  $-$12.8    ($-$14.7,\,$-$10.9) \\ 
 Gender: male  &  941.1    (883.5,\,998.7) \\ 
 Ethnicity: white  &  17.7    ($-$42.4,\,77.7) \\ 
 Hypertensive  &  $-$2.0    ($-$67.5,\,63.5) \\ 
 Education: Less than 9th grade  & 0 (reference)~~ \\
 Education: 9-11th grade  &  272.5    (145.9,\,399.1) \\ 
 Education: High school graduate  &  282.5    (166.7,\,398.3) \\ 
 Education: Some college or AA degree  &  320.8   (207.8,\,433.7) \\ 
 Education: College graduate  &  371.1    (255.6,\,486.7) \\ 
 &  \\ 
\multicolumn{2}{l}{\textbf{The model for the change in systolic blood pressure} $\Delta \E(Y \given \doo(Z),W)$}  \\
&  \\ 
 Parameter & Estimate (95\% CI) \\
\hline
 Mean age (years)  &  $-$0.076    ($-$0.126,\,$-$0.025) \\ 
 Gender: male (\%)  &  0.008    ($-$0.027,\,0.043) \\ 
 Ethnicity: white (\%)  &  0.039    (0.017,\,0.061)~~ \\ 
 Normotensive: Change in UNa (mmol/24h)  &  0.046    (0.017,\,0.075)~~ \\ 
 Hypertensive: Change in UNa (mmol/24h)  &  0.069    (0.040,\,0.098)~~ 
\end{tabular} 
\end{center}
\end{table}

The models are combined according to formula~\eref{eq:realdataE} using the NHANES sampling weights in the averaging. It is assumed here that the urinary sodium and the sodium intake measured in NHANES correspond to each other. The estimated average changes in the systolic blood pressure in the whole population and some subgroups are given in Table~\ref{tab:realdata_results}.  The confidence intervals for the combined results are calculated by applying non-parametric bootstrap  \citepp{diciccio1996bootstrap} simultaneously for both data sources. According to the results, a regular salt user (salt score 6) with hypertension could reduce his or her sodium intake by \SI{7.9}{mmol} (\SI{0.46}{g}, equals \SI{1.2}{g} of salt) and systolic blood pressure by \SI{3.0}{mm Hg} on average by discontinuing the use of salt in preparation and at table. 

\begin{table} 
\caption{The estimated average changes in the systolic blood pressure (in mm Hg) if the use of salt in preparation and at table is discontinued. The weighted average treatment effect (WATE) with NHANES weights is estimated for the whole population as well as subgroups that prefer adding salt or are hypertensive. The results are obtained by combining a meta-analysis of RCTs and NHANES 2013--2016 survey data. \label{tab:realdata_results}}
\begin{center}
\begin{tabular}{ll}
Quantity & Estimate (95\% CI) \\
\hline
 WATE  &  $-$1.2    ($-$4.3,0.5) \\ 
 WATE for salt score 4--6  &  $-$1.5  ($-$4.0,\,0.1) \\ 
 WATE for salt score 6  &  $-$1.9    ($-$4.2,\,$-$0.4) \\ 
 WATE for hypertensive  &  $-$2.0    ($-$5.8,\,0.3) \\ 
 WATE for hypertensive with salt score 4--6  &  $-$2.2    ($-$5.5,\,$-$0.3) \\ 
 WATE for hypertensive with salt score 6  &  $-$3.0    ($-$6.2,\,$-$0.9)  
\end{tabular}
\end{center}
\end{table}

\subsection{Scenarios with Selection Bias and Missing Data} \label{sec:missingdata}
\begin{figure}
\centering
\begin{subfigure}{0.475\textwidth}
 \centering
\begin{tikzpicture}[scale=2.0]
\node [dot = {-0.05}{0}{X}{below}] at (0,0) {};
\node [dot = {0}{0}{Z}{below}] at (1,0) {};
\node [dot = {0.05}{0}{Y}{below}] at (2,0) {};
\node [dot = {0}{0}{R_X}{below}] at (0,0.618) {};
\node [dot = {0}{-0.05}{R_Z}{below}] at (1,0.618) {};
\node [dot = {0}{0}{R_Y}{below}] at (2,0.618) {};

\draw [->] (X) -- (Z);
\draw [->] (Z) -- (Y);
\draw [->] (X) -- (R_Z);
\draw [->] (Y) -- (R_Z);
\draw [<->,dashed] (X) to [bend right=35]  (Y);
\draw [<->,dashed] (R_X) to [bend left=35]  (R_Y);
\draw [<->,dashed] (R_X) to  [bend left=5]  (R_Z);
\draw [<->,dashed] (R_Z) to  [bend left=5] (R_Y);
\end{tikzpicture} 
\caption{Selective sampling (depending on $X$ and $Y$) for mediator $Z$. } \label{fig:Z_MARXY}
\end{subfigure}
\hfill
\begin{subfigure}{0.475\textwidth}
 \centering
\begin{tikzpicture}[scale=2.0]
\node [dot = {-0.05}{0}{X}{left}] at (0,0) {};
\node [dot = {0}{0}{Z}{below}] at (1,0) {};
\node [dot = {0.05}{0}{Y}{right}] at (2,0) {};
\node [dot = {0}{0}{R_X}{left}] at (0,0.618) {};
\node [dot = {0}{0}{R_Z}{above}] at (1,0.618) {};
\node [dot = {0}{0}{R_Y}{right}] at (2,0.4) {};

\draw [->] (X) -- (Z);
\draw [->] (Z) -- (Y);
\draw [->] (Y) -- (R_Y);
\draw [->] (R_Y) -- (R_X);
\draw [->] (R_Y) -- (R_Z);
\draw [<->,dashed] (X) to [bend right=35]  (Y);
\draw [<->,dashed] (R_X) to [bend left=45]  (R_Y);
\draw [<->,dashed] (R_X) to  [bend left=15]  (R_Z);
\draw [<->,dashed] (R_Z) to  [bend left=30] (R_Y);
\end{tikzpicture} 
\caption{Case-control design: selective sampling (depending on $Y$) for $X$, $Z$ and $Y$.} \label{fig:casecontrol}
\end{subfigure}
\caption{Causal models for the front-door setting with missing data. } \label{fig:missing}
\end{figure}
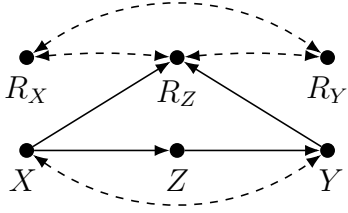
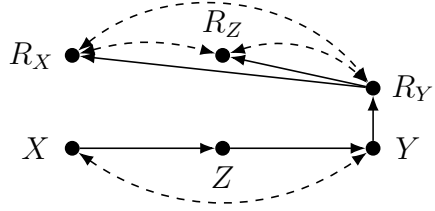

Scenarios where some data are missing by design or unintentionally can be analyzed with \texttt{do-search} as well. For instance, the decision to measure $Z$ may depend on the measurements for $X$ and $Y$. In our example, this could mean that salt intake is measured only for a subgroup where individuals with exceptionally low or high blood pressure are overrepresented. In addition, variables $X$ and $Y$ may suffer from occasional missing values. The graph for this scenario is presented in Figure~\ref{fig:Z_MARXY}. Variables $R_X$, $R_Z$ and $R_Y$ are response indicators for $X$, $Z$ and $Y$, respectively (Section~\ref{sec:dosearch}). The observed data contain variables $X^*$, $Z^*$ and $Y^*$, which are defined as in equation~\eref{eq:missingness}.
A shortcut notation $R_{XZY}=1$ is used to denote $R_X = 1, R_Z = 1, R_Y = 1$.

The causal effect $P(Y \cond \doo(X))$ is identifiable under the assumptions encoded in the graph of Figure \ref{fig:Z_MARXY}  if information on $P(X^*,Y^*,Z^*,R_X,R_Y,R_Z)$ is available. If $Z$ is missing by design, we also know $P(R_Z \given \doo(X,Y))$ which is not necessary for identification. The formula obtained by \texttt{do-search} for the causal effect in Figure~\ref{fig:Z_MARXY}  can be presented as follows 
\begin{equation*}
\begin{aligned}
& P(Y \cond \doo(X)) = \sum_{Z} \bigg[
\sum_{Y^\prime} P(Z \cond X,Y^\prime, R_{XZY}=1)  P(Y^\prime \cond X,R_{XY} = 1) \times
 \\
& \quad \sum_{X^\prime}  P(X^\prime \cond R_{XY} = 1) 
 \frac{P(Z \cond X^\prime,Y,R_{XZY}=1) P(X^\prime,Y,R_{XY} = 1)}{\sum_{Y^\prime} P(Z \cond X^\prime,Y^\prime, R_{XZY}=1) P(X^\prime,Y^\prime, R_{XY}=1)}
\bigg].
\end{aligned}
\end{equation*}

The graph in Figure~\ref{fig:casecontrol} represents a case-control design in the front-door setting\citepp{Karvanen2015studydesign,dosearch}. The selection to the study depends on $Y$. The causal effect $P(Y \cond \doo(Z))$ is not directly identifiable from $P(X^*,Y^*,Z^*,R_X,R_Y,R_Z)$. 
Additional knowledge on the population distribution $P(Y)$ or on the selection mechanism $P(R_Y \cond Y)$ enables \texttt{do-search} to identify the effect. In both cases, a formula for the causal effect $P(Y \given \doo(X))$ can be presented as
\begin{equation*}
\begin{aligned}
&\quad \sum_Z \left[ \frac{\sum_{Y^\prime} P(Y^\prime)P(X,Z \given Y^\prime,R_{XZY}=1)}{\sum_{Z^\prime,Y^\prime} P(Y^\prime)P(X,Z^\prime \given Y^\prime,R_{XZY}=1)}\, \right. \times \\
&\qquad \sum_{X^\prime} \left( \left( \sum_{Y^\prime} P(Y^\prime)P(X^\prime\given Y^\prime,R_{XZY}=1) \right) \, \right. 
\; \left. \left. \frac{P(Y)P(X^\prime,Z \given Y, R_{XZY}=1)}{\sum_{Y^\prime} P(Y^\prime)P(X^\prime,Z \given Y^\prime,R_{XZY}=1)} \vphantom{\sum_{Z^\prime}}\right) \right].
\end{aligned}
\end{equation*}

\section{Discussion}

\texttt{Do-search} offers an effortless way to check identifiability of causal effects. Automated processing saves the time of researchers for tasks where the expert knowledge is necessary. We are not aware of other tools that can, for instance, determine identifiability from an arbitrary chain of experiments 
(Sections~\ref{sec:frontdoor} and \ref{sec:variants}) or solve complicated missing data problems (Section~\ref{sec:missingdata}). \texttt{Do-search} is useful in both planning and analysis of studies. In research design, \texttt{do-search} can be used to check whether the new data to be collected will enable the effect of interest to be identified, or to determine whether an existing dataset will be beneficial in a secondary analysis when combined with other data sources. Although not considered in this paper, \texttt{do-search} can also handle selection diagrams\citepp{bareinboim2015recovering,dosearch} and solve transportability problems\citepp{bareinboim2014transportability,dosearch} where the data sources may originate from heterogeneous domains. 

The key structural assumptions used by \texttt{do-search} are encoded as a causal graph, which can be freely specified by the researcher. 
This specification requires epidemiological expertise on the subject of interest. In addition to causal relations, the graph has to specify the understanding on the selection and missing data mechanism. The input data sources are given in easily interpretable symbolic non-parametric form. 

There are important issues that are outside the scope of \texttt{do-search}. The researcher has to evaluate the quality of the data in each study considered. Especially, the researcher has to decide how to consider variables that aim to measure the same underlying phenomenon but have different definitions. For example, survey questions ``Did you add salt at table yesterday?'' and ``Do you usually add salt at table?'' both aim to measure salt-adding behavior but may have differing answers for the same individual. 

A theoretical limitation of \texttt{do-search} is that the current set of rules used are not yet complete for all missing data problems. Another limitation related to incomplete data is that the missing data mechanisms may not differ between the data sources.  
 The scalability of \texttt{do-search} may be an issue for some applications,
  because the computational complexity increases exponentially as the number of variables increases. However, in many cases, it is possible to reduce the computational burden by grouping variables together that have a similar role in the causal model, i.e. having $Z$ to represent a group of variables instead of a single variable. 

\texttt{Do-search} operates fully non-parametrically and cannot utilize parametric assumptions that could extend identifiability as such.
For instance, if the functional relations are linear, instrumental variables may render non-parametrically non-identifiable effects  identifiable \citepp{angrist1996,chen2017}.
Recently, \texttt{do-search} has been extended to make use of a particular type of restrictions, known as context-specific independence relations \citepp{tikka2019csi}.
However, non-parametric non-identifiability and parametric identifiability indicate that the estimation results may be highly sensitive to the parametric assumptions made.

The causal effect of salt-adding behavior on blood pressure was estimated by combining observational and experimental data. The obtained estimates are not directly comparable with earlier results but are well aligned with some related studies when the uncertainty is taken into account\citepp{takahashi2006blood,he2009comprehensive,kelly2016effect}. It is known that the sodium intake measured by a dietary questionnaire is prone for recall bias and thus less reliable than the sodium intake measured from urine \citepp{karppanen1998sodium,he2009comprehensive}. In our results, the relatively wide confidence intervals of the point estimates may reflect the measurement error of salt intake in NHANES. The analysis could be continued further, and questions such as the causal effect of salt-adding behavior on the risk of cardiovascular diseases and total mortality  \citepp{aburto2013effect,alderman2012dietary} could be studied in a similar manner.

\texttt{Do-search} is a versatile addition to the epidemiological toolbox. Combined with other tools it paths the way towards a holistic approach that goes beyond the traditional meta-analysis and allows for a systematic analysis of cumulative evidence from heterogeneous data sources.

\section*{Acknowledgements}
ST was supported by Academy of Finland grant 311877 (Decision analytics utilizing causal models
and multiobjective optimization). AH was supported by Academy of Finland grant 295673. The authors thank Jukka Nyblom for useful comments.

\bibliographystyle{apalike}
\bibliography{bibliography}

\newpage
\section*{Appendix A: R code for the examples}

\begin{small}
\begin{verbatim}
library(dosearch)

#####################
# Graphs of Figure 1

graph1a <- "
    X -> Z
    Z -> Y
    Z -> Y
    X <-> Y
  "

graph1b <- "
    X -> Z
    Z -> Y
    X <-> Z
"

graph1c <- "
    X -> Z
    Z -> Y
    X <-> Y
    X <-> Z
  "

graph1d <- "
    X -> Z
    Z -> Y
    Z <-> Y
  "

graph1e <- "
    X -> Z
    Z -> Y
    Z <-> Y
    X <-> Y
  "
graph1f <- "
    X -> Z
    Z -> Y
    W -> X
    W -> Z
    W -> Y
    X <-> Y
  "

graph1g <- "
    X -> Z
    Z -> Y
    W -> X
    W -> Z
    W -> Y
    X <-> Y
    X <-> Z
  "

graphs <- c(graph1a, graph1b, graph1c, graph1d, graph1e, graph1f, graph1g)

datasources <- c(
  "P(X,Y,Z)",
  "P(X,Z)
   P(Y|do(Z))",
  "P(Z|do(X))
   P(Y|do(Z))",
  "P(Z,Y)
   P(Z|do(X))",
  "P(X,Z)
   P(X,Y)
   P(Z,Y)
   P(Z|do(X))",
  "P(X,Y,Z,W)",
  "P(X,Z,W)
   P(Y|do(Z),W)",
  "P(Z|do(X),W)
   P(Y|do(Z),W)",
  "P(Z|do(X),W)
   P(Y|do(Z),W)
   P(W)"
)
query <- "P(Y|do(X))"

n <- length(datasources)
m <- length(graphs)

id <- matrix("NA", n, m)
formula <- matrix("", n, m)

for(i in 1:n) {
  for(j in 1:m) {
    result <- dosearch(datasources[i], query, graphs[j])
    id[i, j] <- ifelse(result$identifiable, "Y", "N")
    if (result$identifiable) formula[i, j] <- result$formula
  }
}


#####################
# Graph of Figure 3

graph3 <- "
    X -> Z
    Z -> Y
    W -> X
    W -> Z
    W -> Y
    H -> X
    H -> Z
    X <-> Y
    H <-> C
  "
datasources3 <- c(
  "P(X,Z,H,W)
  P(Y|do(Z),W)"
)

query3 <- "P(Y|do(X))"
result3 <- dosearch(datasources3, query3, graph3)

#####################
# Graphs of Figure 4

graph4a <- "
    X -> Z
    Z -> Y
    Z -> Y
    X <-> Y
    X -> R_Z
    Y -> R_Z
    R_X <-> R_Z
    R_X <-> R_Y
    R_Z <-> R_Y
  "

graph4b <- "
    X -> Z
    Z -> Y
    Z -> Y
    X <-> Y
    Y -> R_Y
    R_Y -> R_X
    R_Y -> R_Z
    R_X <-> R_Z
    R_X <-> R_Y
    R_Z <-> R_Y
  "

graphs4 <- c(graph4a, graph4b)

datasources4 <- c(
  "P(X*,Y*,Z*,R_X,R_Y,R_Z)", 
  "P(X*,Y*,Z*,R_X,R_Y,R_Z)
   P(Y)",
  "P(X*,Y*,Z*,R_X,R_Y,R_Z)
   P(R_Y|Y)"
)
mdxyz <- "R_X : X, R_Y : Y, R_Z : Z"
mdxz <- "R_X : X, R_Z : Z"
md <- c(mdxyz,mdxz,mdxyz,mdxyz)

n4 <- length(datasources4)
m4 <- length(graphs4)

id4 <- matrix("NA", n4, m4)
formula4 <- matrix("", n4, m4)

for(i in 1:n4) {
  for(j in 1:m4) {
    cat(i,j,"\n")
    result <- dosearch(datasources4[i], query, graphs4[j], missing_data = md[i])
    id4[i, j] <- ifelse(result$identifiable, "Y", "N")
    if (result$identifiable) formula4[i, j] <- result$formula
  }
}
\end{verbatim}
\end{small}

\section*{Appendix B: Proofs}

Here we provide proofs that the scenarios marked as non-identifiable in Table~1 really are non-identifiable. According to the definition, a causal effect is non-identifiable if there exists two models , $M^1$ and $M^2$, that share the input distributions but differ by the causal effect of interest.



\paragraph{(1)} Data sources: $P(X,Z)$, $P(Y \cond \doo(Z))$. For Figures 1(b), 1(c), 1(f) and 1(g), we define:
\[
M^1 = \begin{cases}
P^1(U = 1) = \frac12 \\
P^1(X = 1 \cond U = 1) = p \\
P^1(X = 1 \cond U = 0) = 1 - p \\
P^1(Z = 1 \cond X = 0,U) = \frac12 \\
P^1(Z = 1 \cond X = 1,U = 1) = q\\
P^1(Z = 1 \cond X = 1,U = 0) = 1-q \\
P^1(Y = 1\cond Z = 1) = a \\
P^1(Y = 1\cond Z = 0) = b
\end{cases}
M^2 = \begin{cases}
P^2(U = 1) = p \\
P^2(X = 1 \cond U) = \frac12 \\
P^2(Z = 1 \cond X = 0,U) = \frac12 \\
P^2(Z = 1 \cond X = 1,U = 1) = q\\
P^2(Z = 1 \cond X = 1,U = 0) = 1-q \\
P^2(Y = 1 \cond Z = 1) = a \\
P^2(Y = 1 \cond Z = 0) = b.
\end{cases}.
\]
It follows that
\[ P^1(X = 0,Z) = P^2(X = 0,Z) = \frac12 \\ \]
\[ P^1(X = 1,Z = 1) = P^2(X = 1,Z = 1) = \frac{qp + (1-q)(1-p)}2 \\ \]
\[ P^1(X = 1,Z = 0) = P^2(X = 1,Z = 0) = \frac{(1-q)p + q(1-p)}2 \\ \]
\[ P^1(Y=1 \cond \doo(Z = 1)) = P^2(Y=1 \cond \doo(Z=1)) = a \]
\[ P^1(Y=1 \cond \doo(Z = 0)) = P^2(Y=1 \cond \doo(Z=0)) = b. \]
but
\begin{align*}
P^1(Y=1 \cond \doo(X = 1)) &= \sum_{Z,U} P^1(Y = 1 \cond Z)P^1(Z \cond X=1,U)P^1(U) \\
 &= aq\frac12 + a(1-q)\frac12 + b(1-q)\frac12 + bq\frac12 \\
 &= \frac{a}2 + \frac{b}2 \\
P^2(Y=1 \cond \doo(X = 1)) &= aqp + a(1-q)(1-p) + b(1-q)p + bq(1-p).
\end{align*}
Thus $P^1(Y=1 \cond \doo(X = 1)) \neq P^2(Y=1 \cond \doo(X = 1))$ when $a \neq b$, $p \neq \frac12$ and $q \neq \frac12$. For Figures 1(d) and 1(e), we define:
\[
M^1 = \begin{cases}
P^1(U = 1) = \frac12 \\
P^1(X = 1) = \frac12 \\
P^1(Z = 1 \cond X, U = 1) = p \\
P^1(Z = 1 \cond X, U = 0) = 1-p \\
P^1(Y = 1 \cond Z=0,U) = \frac12 \\
P^1(Y = 1 \cond Z=1,U = 1) = a \\
P^1(Y = 1 \cond Z=1,U = 0) = b
\end{cases}
M^2 = \begin{cases}
P^2(U = 1) = p \\
P^2(X = 1) = \frac12 \\
P^2(Z = 1 \cond X, U) = \frac12 \\
P^2(Y = 1 \cond Z=0,U) = \frac12. \\
P^2(Y = 1 \cond Z=1,U) = \frac{a + b}2. \\
\end{cases}
\]
It follows that
\[
P^1(X,Z) = P^2(X,Z) = \frac14.
\]
\begin{align*}
P^1(Y = 1 \cond \doo(Z=1)) &= \sum_{X,U} P^1(Y = 1 \cond Z=1,U)P^1(X)P^1(U) \\
 &= \sum_{U} P^1(Y = 1 \cond Z=1,U)P^1(U) \\
 &= a\frac12 + b\frac12 \\
 &= \frac{a+b}2 p + \frac{a+b}2 (1-p) \\
 &= P^2(Y = 1 \cond \doo(Z=1))
\end{align*}
and
\[
P^1(Y = 1 \cond \doo(Z=0)) = P^1(Y = 1 \cond \doo(Z=0)) = \frac12,
\]
but
\begin{align*}
P^1(Y=1 \cond \doo(X = 1)) &= \sum_{Z,U} P^1(Y = 1 \cond Z,U)P^1(Z \cond X = 1,U)P^1(U) \\
 &= ap\frac12 + b(1-p)\frac12 + \frac12(1-p)\frac12 + \frac12p\frac12 \\
 &= \frac{ap}2 + \frac{b(1-p)}2 + \frac14 \\
P^2(Y=1 \cond \doo(X = 1)) &= \frac{a + b}2 \frac12 p + \frac{a + b}2\frac12(1-p) + \frac12 \frac12p + \frac12 \frac12 (1-p) \\
 &= \frac{a + b}4 + \frac14.
\end{align*}
Thus $P^1(Y=1 \cond \doo(X = 1)) \neq P^2(Y=1 \cond \doo(X = 1))$ when $a \neq b$ and $p \neq \frac12$.

\paragraph{(2)} Data sources: $P(Z \cond \doo(X)), P(Y \cond \doo(Z))$. For Figures 1(d), 1(e), 1(f) and 1(g), we define $M^1$ and $M^2$ as in the second construction of (1). This construction also guarantees that $P^1(Z \cond \doo(X)) = P^2(Z \cond \doo(X))$, since:
\begin{align*}
P^1(Z = 1 \cond \doo(X)) &= \sum_U P^1(Z = 1 \cond X, U)P^1(U) \\
 &= p\frac12 + (1-p)\frac12 \\
 &= P^2(Z = 1 \cond \doo(X))
\end{align*}

\paragraph{(3)} Data sources: $P(Z,Y), P(Z \cond \doo(X))$. For Figures 1(a), 1(c), 1(e), 1(f) and 1(g), we define:
\[
M^1 = \begin{cases}
P^1(U = 1) = \frac12 \\
P^1(X = 1 \cond U = 1) = \frac12 \\
P^1(X = 1 \cond U = 0) = \frac14 \\
P^1(Z = 1 \cond X = 1) = \frac34 \\
P^1(Z = 1 \cond X = 0) = \frac14 \\
P^1(Y = 1 \cond Z = 1, U = 1) = \frac45 \\
P^1(Y = 1 \cond Z = 1, U = 0) = \frac{7}{10} \\
P^1(Y = 1 \cond Z = 0, U = 1) = \frac{13}{20} \\
P^1(Y = 1 \cond Z = 0, U = 0) = \frac{1}{20}
\end{cases}
M^2 = \begin{cases}
P^2(U = 1) = \frac12 \\
P^2(X = 1 \cond U = 1) = \frac12 \\
P^2(X = 1 \cond U = 0) = \frac14 \\
P^2(Z = 1 \cond X = 1) = \frac34 \\
P^2(Z = 1 \cond X = 0) = \frac14 \\
P^2(Y = 1 \cond Z = 1,U = 1) = \frac{19}{20} \\
P^2(Y = 1 \cond Z = 1,U = 0) = \frac12 \\
P^2(Y = 1 \cond Z = 0,U = 1) = \frac{2}{5} \\
P^2(Y = 1 \cond Z = 0,U = 0) = \frac14
\end{cases}
\]
It follows that
\begin{align*}
P^1(Y=1,Z=1) &= P^2(Y=1,Z=1) = \frac{53}{160} \\
P^1(Y=1,Z=0) &= P^2(Y=1,Z=0) = \frac{57}{320} \\
P^1(Y=0,Z=1) &= P^2(Y=0,Z=1) = \frac{17}{160} \\
P^1(Y=0,Z=0) &= P^2(Y=0,Z=0) = \frac{123}{320}
\end{align*}
and
\begin{align*}
P^1(Z = 1 \cond \doo(X = 1)) &= P^2(Z = 1 \cond \doo(X = 1)) = \frac34 \\
P^1(Z = 1 \cond \doo(X = 0)) &= P^2(Z = 1 \cond \doo(X = 0)) = \frac14,
\end{align*}
but
\[
P^1(Y=1 \cond \doo(X = 1)) = \frac{13}{20} \neq \frac58 = P^2(Y=1 \cond \doo(X = 1)).
\]
For Figure~1(d), we define:
\[
M^1 = \begin{cases}
  P^1(U = 1) = \frac12 \\
  P^1(X = 1) = \frac25 \\
  P^1(Z = 1 \cond X = 1, U = 1) = \frac25 \\
  P^1(Z = 1 \cond X = 1, U = 0) = \frac{7}{20} \\
  P^1(Z = 1 \cond X = 0, U = 1) = \frac{3}{10} \\
  P^1(Z = 1 \cond X = 0, U = 0) = \frac25 \\
  P^1(Y = 1 \cond Z = 1, U = 1) = \frac15\\
  P^1(Y = 1 \cond Z = 1, U = 0) = \frac{3}{10}\\
  P^1(Y = 1 \cond Z = 0, U = 1) = \frac{3}{10}\\
  P^1(Y = 1 \cond Z = 0, U = 0) = \frac{7}{20}
\end{cases}
M^2 = \begin{cases}
  P^2(U = 1) = \frac12 \\
  P^2(X = 1) = \frac25 \\
  P^2(Z = 1 \cond X = 1, U = 1) = \frac{1}{10} \\
  P^2(Z = 1 \cond X = 1, U = 0) = \frac{7}{20}  \\
  P^2(Z = 1 \cond X = 0, U = 1) = \frac12 \\
  P^2(Z = 1 \cond X = 0, U = 0) = \frac15 \\
  P^2(Y = 1 \cond Z = 1, U = 1) = \frac15 \\
  P^2(Y = 1 \cond Z = 1, U = 0) = \frac{3}{10}\\
  P^2(Y = 1 \cond Z = 0, U = 1) = \frac{3}{10}\\
  P^2(Y = 1 \cond Z = 0, U = 0) = \frac{7}{20}
\end{cases}
\]
It follows that
\begin{align*}
P^1(Y=1,Z=1) &= P^2(Y=1,Z=1) = \frac{91}{1000} \\
P^1(Y=1,Z=0) &= P^2(Y=1,Z=0) = \frac{601}{2000} \\
P^1(Y=0,Z=1) &= P^2(Y=0,Z=1) = \frac{269}{1000} \\
P^1(Y=0,Z=0) &= P^2(Y=0,Z=0) = \frac{679}{1000}
\end{align*}
and
\begin{align*}
P^1(Z = 1 \cond \doo(X = 1)) &= P^2(Z = 1 \cond \doo(X = 1)) = \frac38 \\
P^1(Z = 1 \cond \doo(X = 0)) &= P^2(Z = 1 \cond \doo(X = 0)) = \frac{7}{20},
\end{align*}
but
\[
P^1(Y=1 \cond \doo(X = 1)) = \frac{63}{160} \neq \frac{57}{160} = P^2(Y=1 \cond \doo(X = 1)).
\]

\paragraph{(4)} Data sources: $P(X,Z), P(X,Y), P(Z,Y), P(Z \cond \doo(X))$. For Figures 1(a), 1(c), 1(e),
1(f) and 1(g), we use the first construction of (3). It remains to show that $P^1(X,Z) = P^2(X,Z)$ and $P^1(X,Y) = P^2(X,Y)$ also hold.
A simple computation gives:
\begin{align*}
P^1(X=1,Y=1) &= P^1(X=1,Y=1) = \frac{33}{128} \\
P^1(X=1,Y=0) &= P^1(X=1,Y=0) = \frac{15}{128} \\
P^1(X=0,Y=1) &= P^1(X=0,Y=1) = \frac{161}{640} \\
P^1(X=0,Y=0) &= P^1(X=0,Y=0) = \frac{239}{640}.
\end{align*}
We know that $P^1(U) = P^2(U)$, $P^1(X \cond U) = P^2(X \cond U)$ and $P^1(Z \cond X) = P^2(Z \cond X)$, which means that $P^1(X,Z) = P^2(X,Z)$ as well.

\paragraph{(5)} Data sources: $P(X,Z,W), P(Y \cond \doo(Z), W)$. For Figures 1(b), 1(c), 1(d), 1(e), the data sources are essentially the same as in (1) since $W$ is unconnected. Thus the constructions of (1) are also applicable here. For Figure 1(g) we define:
\[
M^1 = \begin{cases}
P^1(U_1 = 1) = \frac12 \\
P^1(U_2 = 1) = \frac12 \\
P^1(W = 1) = \frac12 \\
P^1(X = 1 \cond W,U_1 = 1,U_2) = p \\
P^1(X = 1 \cond W,U_1 = 0,U_2) = 1 - p \\
P^1(Z = 1 \cond W,X = 0,U_1) = \frac12 \\
P^1(Z = 1 \cond W,X = 1,U_1 = 1) = q\\
P^1(Z = 1 \cond W,X = 1,U_1 = 0) = 1-q \\
P^1(Y = 1 \cond W,Z = 1,U_2) = a \\
P^1(Y = 1 \cond W,Z = 0,U_2) = b
\end{cases}
\!M^2 = \begin{cases}
P^2(U_1 = 1) = p \\
P^2(U_2 = 1) = \frac12 \\
P^2(W = 1) = \frac12 \\
P^2(X = 1 \cond W,U_1,U_2) = \frac12 \\
P^2(Z = 1 \cond W,X = 0,U_1) = \frac12 \\
P^2(Z = 1 \cond W,X = 1,U_1 = 1) = q\\
P^2(Z = 1 \cond W,X = 1,U_1 = 0) = 1-q \\
P^2(Y = 1 \cond W,Z = 1,U_2) = a \\
P^2(Y = 1 \cond W,Z = 0,U_2) = b.
\end{cases}
\]
The parametrization is very similar to (1) and it follows that
\[ P^1(X = 0,Z,W) = P^2(X = 0,Z,W) = \frac18 \]
\[ P^1(X = 1,Z = 1,W) = P^2(X = 1,Z = 1,W) = \frac{qp + (1-q)(1-p)}4 \]
\[ P^1(X = 1,Z = 0,W) = P^2(X = 1,Z = 0,W) = \frac{(1-q)p + q(1-p)}4 \]
\[ P^1(Y=1 \cond \doo(Z = 1),W) = P^2(Y=1 \cond \doo(Z=1),W) = a \]
\[ P^1(Y=1 \cond \doo(Z = 0),W) = P^2(Y=1 \cond \doo(Z=0),W) = b. \]
but
\begin{align*}
P^1(Y=1 \cond \doo(X = 1)) &= \sum_{Z,W,U_1,U_2} P^1(Y = 1 \cond W,Z,U_2)P^1(Z \cond X=1,W,U_1)P^1(W)P^1(U_1)P^1(U_2) \\
&= \frac18 \sum_{Z,W,U_1,U_2} P^1(Y = 1 \cond Z)P^1(Z \cond X=1,W,U_1) \\
 &= \frac12\left(aq + a(1-q) + b(1-q) + bq\right) \\
 &= \frac{a}2 + \frac{b}2 \\
P^2(Y=1 \cond \doo(X = 1)) &= aqp + a(1-q)(1-p) + b(1-q)p + bq(1-p).
\end{align*}
Thus $P^1(Y=1 \cond \doo(X = 1)) \neq P^2(Y=1 \cond \doo(X = 1))$ when $a \neq b$, $p \neq \frac12$ and $q \neq \frac12$.

\paragraph{(6)} Data sources: $P(Z \cond \doo(X), W), P(Y \cond \doo(Z), W)$. For Figures 1(d) and 1(e), the data sources are essentially the same as in (2) since $W$ is unconnected. Thus the construction of (2) is also applicable here. For Figures 1(f) (assume that $U_2$ is unconnected) and 1(g) we define:
\[
M^1 = \begin{cases}
P^1(U_1 = 1) = \frac12 \\
P^1(U_2 = 1) = \frac12 \\
P^1(W = 1) = \frac12 \\
P^1(X = 1 \cond W = 1,U_1,U_2) = p \\
P^1(X = 1 \cond W = 0,U_1,U_2) = 1 - p \\
P^1(Z = 1 \cond X = 0,W,U_1) = \frac12 \\
P^1(Z = 1 \cond X = 1,W = 1,U_1) = q\\
P^1(Z = 1 \cond X = 1,W = 0,U_1) = 1-q \\
P^1(Y = 1 \cond Z = 1,W,U_2) = a \\
P^1(Y = 1 \cond Z = 0,W,U_2) = b
\end{cases}
\!M^2 = \begin{cases}
P^2(U_1 = 1) = \frac12 \\
P^2(U_2 = 1) = \frac12 \\
P^2(W = 1) = p \\
P^2(X = 1 \cond W,U_1,U_2) = \frac12 \\
P^2(Z = 1 \cond X = 0,W,U_1) = \frac12 \\
P^2(Z = 1 \cond X = 1,W = 1,U_1) = q\\
P^2(Z = 1 \cond X = 1,W = 0,U_1) = 1-q \\
P^2(Y = 1 \cond Z = 1,W,U_2) = a \\
P^2(Y = 1 \cond Z = 0,W,U_2) = b.
\end{cases}
\]
We have that
\begin{align*}
P^1(Z \cond \doo(X = 0),W) &= P^2(Z \cond \doo(X = 0),W) = \frac12 \\
P^1(Z \cond \doo(X = 1),W=1) &= P^2(Z \cond \doo(X = 1),W=1) = q \\
P^1(Z \cond \doo(X = 1),W=0) &= P^2(Z \cond \doo(X = 1),W=0) = 1-q \\
P^1(Y \cond \doo(Z = 1),W) &= P^2(Z \cond \doo(Z = 1),W) = a \\
P^1(Y \cond \doo(Z = 0),W) &= P^2(Z \cond \doo(Z = 0),W) = b.
\end{align*}
However
\begin{align*}
P^1(Y=1 \cond \doo(X = 1)) &= \sum_{Z,W,U_1,U_2} P^1(Y = 1 \cond W,Z,U_2)P^1(Z \cond X=1,W,U_1)P^1(W)P^1(U_1)P^1(U_2) \\
&= \frac18 \sum_{Z,W,U_1,U_2} P^1(Y = 1 \cond Z)P^1(Z \cond X=1,W,U_1) \\
 &= \frac12\left(aq + a(1-q) + b(1-q) + bq\right) \\
 &= \frac{a}2 + \frac{b}2 \\
P^2(Y=1 \cond \doo(X = 1)) &= aqp + a(1-q)(1-p) + b(1-q)p + bq(1-p).
\end{align*}
Thus $P^1(Y=1 \cond \doo(X = 1)) \neq P^2(Y=1 \cond \doo(X = 1))$ when $a \neq b$, $p \neq \frac12$ and $q \neq \frac12$.

\paragraph{(7)} Data sources: $P(Z \cond \doo(X), W), P(Y \cond \doo(Z), W), P(W)$. For Figures 1(d) and 1(e), the data sources are essentially the same as in (6) and (2) since $W$ is unconnected. Thus the construction of (2) is also applicable here.

\end{document}